\documentclass{mem}
\usepackage{natbib}\usepackage{txfonts}\usepackage{balance}
\usepackage{graphicx}
\usepackage[a4paper]{hyperref}
\idline{00}{00}
\begin{document}

\title{X-ray emission from early-type galaxies}

   \subtitle{}

\author{S. \,Pellegrini\inst{1} 
          }

  \offprints{S. Pellegrini}

\institute{
Dipartimento di Astronomia, Universit\`a di Bologna, 
via Ranzani 1, I-40127 Bologna, Italy
\email{silvia.pellegrini@unibo.it}
}

\authorrunning{Pellegrini}

\titlerunning{X-rays from E/S0s}

\abstract{
  The last $\sim$10 years have seen a large progress in the X-ray
  investigation of early-type galaxies of the local universe, and
  first attempts have been made to reach redshifts $z>0$ for these
  objects, thanks to the high angular resolution and sensitivity of
  the satellites $Chandra$ and $XMM-$Newton.  Major advances have been
  obtained in our knowledge of the three separate contributors to the
  X-ray emission, that are the stellar sources, the hot gas and the
  galactic nucleus. Here a brief outline of the main results is
  presented, pointing out the questions that remain open, and finally
  discussing the prospects to solve them with a wide area X-ray survey
  mission such as $WFXT$.

}
\maketitle{}

\section{Introduction} 

X-ray investigations of early-type galaxies\footnote{This work is
devoted to "normal" early-type galaxies, where the X-ray emission is
not dominated by an AGN, and keeps below $\sim 10^{42}$ erg s$^{-1}$.}
(hereafter ETGs) of the local universe began in the 1980s with the
{\it Einstein} satellite, and revealed that the total X-ray luminosity
originates from a combination of hot interstellar gas and low-mass
X-ray binaries \citep[LMXBs;][]{fab89}. With the advent of the
$ROSAT$, $ASCA$ and then $Chandra$ and XMM-Newton eras, our knowledge
of all the components of the X-ray emission has deepened considerably:
among stellar sources by far the largest contribution comes from
LMXBs, and it has been quantified; a hot gaseous halo (with a
temperature of $\sim $few million degrees) can be present with largely
varying amounts; another important galactic component, a supermassive
black hole (MBH) believed to be common at the center of ETGs and a
relic of the past quasar activity, showed luminosities ranging
continuously from the lowest detectable levels (e.g., that of a bright
LMXB in Virgo) to values typical of Seyferts. The combined study of
the hot gas and low luminosity nuclei turned out to be a crucial tool
to build our understanding of MBH accretion and feedback in the local
universe.

\begin{figure*}[t!]
\resizebox{\hsize}{!}{\includegraphics[clip=true]{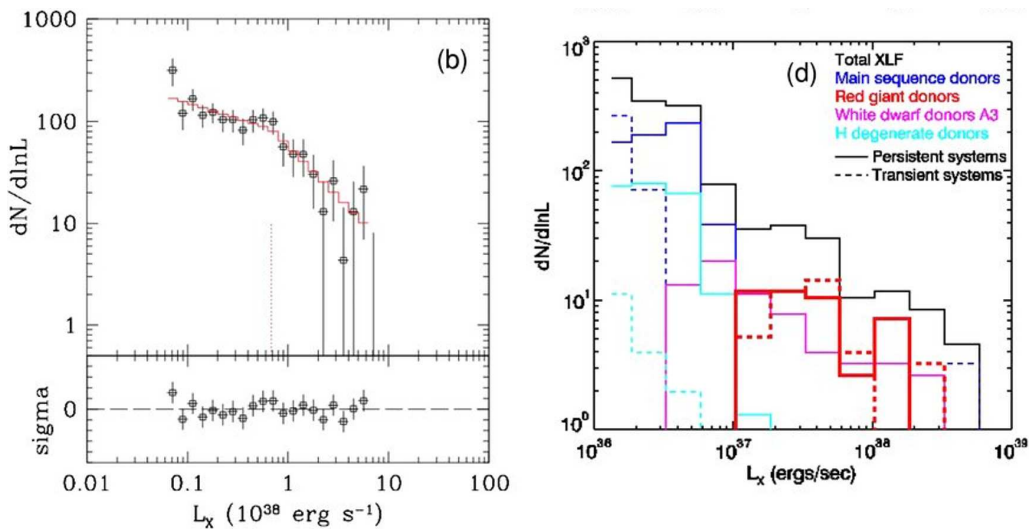}}
\caption{\footnotesize
The 0.3--8 keV luminosity function of LMXBs for three hot gas poor ETGs
with deep $Chandra$ pointings
(Sect.~\ref{dscr}) on the left, and the theoretical prediction 
\cite{fra08} on the right \citep[from][]{kim09}.
}
\label{f1}
\end{figure*}


The results above are based on few tens of ETGs accurately studied
with $Chandra$ and $XMM$-Newton, whose archives contain at present
roughly two hundreds ETGs with a specific pointing, located within a
distance of $\sim 100$ Mpc.  In this work I review briefly the main
advances concerning the three major emission components (the
stellar emission in Sect.~\ref{dscr}, the hot gas in Sect.~\ref{ism}
and the galactic nuclei in Sect.~\ref{nuc}), indicating also the needs for
further investigation; in Sect.~\ref{hiz} I summarize the current
scant and sparse knowledge of the X-ray properties of ETGs beyond the
local universe; in Sect.~\ref{wfxt} I discuss the prospects to address
a few important science goals with $WFXT$.

\section{Stellar sources}\label{dscr}

The stellar X-ray emission of ETGs is contributed by a population of
weak sources ($L_X<10^{34}$ erg s$^{-1}$) as late type stellar
coronae, cataclismic variables, and coronally active binaries
\citep{pf94}, and by the more luminous LMXBs, associated with an old
stellar population and powered by accretion from a low-mass late-type
star onto a compact stellar remnant, a neutron star or a black hole.
The origin and evolution of the collective LMXB population of ETGs is
the subject of much discussion \citep{fab06}; LMXBs are found in both
the stellar field and globular clusters, but their incidence per unit
stellar mass is much higher in the latter, suggesting the importance
of a dynamical formation mechanism.

Exploiting the sub-arcsecond angular resolution provided by $Chandra$
the nature of the stellar contribution to the X-ray emission could be
better constrained, especially with deep pointings at ETGs (almost)
devoid of an important contaminant such as the hot gas
\citep[2009]{bra08}.  In this way the collective contribution of the
weak population could be estimated in NGC3379 \citep{rev}. Luminous
($L_X>10^{36}$ erg s$^{-1}$) pointlike sources could instead be
individually detected and their X-ray luminosity function (XLF) be
built in a number of galaxies, the deepest studies being those for
NGC3379, NGC4278 and NGC4697 \citep[][see Fig.~\ref{f1}] {kim09}, and
NGC5128 \citep{vos}.  One major goal is to calibrate the dependence of
the collective X-ray emission from LMXBs on the galaxy stellar mass or
luminosity, age and globular cluster specific frequency.  The high
luminosity end of the XLF ($L_X>$several$\times 10^{37}$ erg s$^{-1}$)
and the collective luminosity of the whole LMXB population as a
function of the galactic luminosity are now reasonably known, with a
possible dependence also on the globular cluster specific frequency
still to be evaluated \citep{kim04,gil,kim09}.  The features in the
observed XLFs that are being discovered (as breaks at high and low
luminosities, possible bumps, differences for field and globular
cluster sources) represent important inputs to theoretical models for
LMXB formation and evolution, as those built with the advanced
population synthesis code StarTrack \citep[][2009]{fra08}.  These
models also predict the evolution of the XLF with galaxy age, and then
the collective (hard) emission from LMXBs; such predictions are useful
for investigations of ETGs at higher redshift that are attempted
currently (Sect.~\ref{hiz}) and will flourish with $WFXT$
(Sect.~\ref{wfxt}).

\section{Hot interstellar medium}\label{ism}

\begin{figure}[t!]
\resizebox{\hsize}{!}{\includegraphics[clip=true]{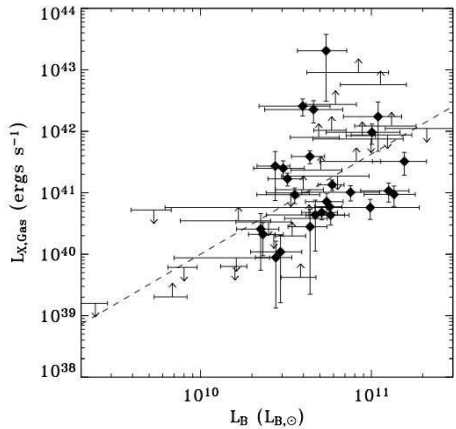}}
\caption{\footnotesize
The 0.3--5 keV luminosity of the hot gas as a function of absolute
blue galactic luminosity, for a sample of ETGs of the local universe
in the $Chandra$ archive \citep[from][]{ds07}.
}
\label{f2}
\end{figure}

$Chandra$ observations allowed to separate the contribution of
stellar sources and hot gas, as well as emission coming from different
spatial regions within galaxies, obtaining the best definition ever
for the hot gas properties \citep[e.g.,][]{kf,hum}. It is now proven that
in optically luminous ETGs the soft interstellar gas can be present
with largely varying amounts \citep{fab89,pel99,sar01}, producing a scatter
in $L_X$ up to a factor of 100 at fixed galactic optical luminosity
(Fig.~\ref{f2}); in optically faint ETGs instead 
the X-ray emission is
always dominated by LMXBs \citep{dav06,pel07,trin}.

The hot ISM provides fuel for the central MBH and absorbs energy from
nuclear outbursts, in a complex cycle whose mechanism is not yet fully
understood \citep[e.g.,][]{for05,bal09,co09}. A compilation of radial
temperature profiles for the hot gas shows that the radio luminosity
decreases continuously as gradients in the profiles change from
positive to negative, as if the profiles were reversing the
temperature gradient over time following an activity cycle
\citep{ds08}.  Also the environment in which ETGs reside can influence
the hot gas coronae, having an effective impact on their outer
temperature gradient \citep{ds08}, and on their size and luminosity
via stripping, sloshing, compression, conduction \citep{sun07}; the
environment is also important for the injection of metals from ETGs in
the intracluster medium \citep[e.g.,][]{kim08}.  The sample of ETGs
for which all these phenomena have been investigated is however
limited, and real samples in a statistical sense (i.e., made of
thousands of objects) are needed to establish clearly the effects of a
surrounding medium, of the interactions with neighbours, of feedback,
possibly dividing galaxies based on mass, age, and kinds of environment.

\section{Low luminosity MBHs}\label{nuc}

\begin{figure}[t!]
\vskip -2.5truecm
\hskip -1truecm
\resizebox{10cm}{!}
{\includegraphics{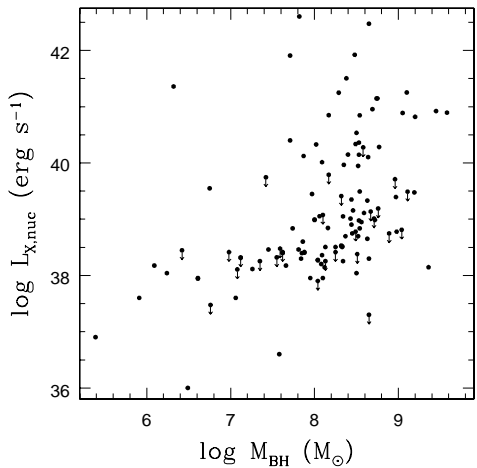}}
\vskip -1truecm
\caption{\footnotesize
The 2--10 keV nuclear luminosity as a function of the MBH mass
for a sample of ETGs of the local universe
in the $Chandra$ archive \citep[from][]{P10}.
}
\label{f3}
\end{figure}

Thanks to $Chandra$'s angular resolution, for the first time
measurement of the nuclear X-ray emission down to values as low as
$10^{39}$ erg s$^{-1}$ and out to distances of $\sim 60$ Mpc were
obtained.  MBHs of the local universe turned out to be typically very
sub-Eddington emitters \citep{pel05,gal08} and their radiative
quiescence was interpreted in terms of radiatively inefficient
accretion \citep[RIAF;][]{nar95}, possibly with the mechanical power
dominating the total energy output of accretion \citep[e.g.,][]{all}.
From the sample available, there appears to be only a weak relation of
the nuclear luminosity with the MBH mass or with the galactic hot gas
content, with a very large dispersion dominating the two relations
\cite{P10}; Fig.~\ref{f3}.  The modeling of the
observables (mainly the nuclear spectral energy distribution from
radio to X-rays, and the mass accretion rate derived from the gas
density and temperature close to the accretion radius) allows to
establish the origin of the nuclear X-rays, from a standard disk plus hot
corona, a RIAF, a jet, or a combination of them \cite[e.g.,][]{fab03,pt04};
all this gives important clues on the modality of the MBH feeding, on the
kinetic feedback, and then on the co-existence of MBHs and
host galaxies.  Despite many efforts applied to observational data,
accretion in the local universe remains poorly known, while its
knowledge is important for a complete understanding of the MBH-host
galaxy coevolution process.  Current believes are that MBHs spend most
of their life in the RIAF regime \citep{hop,co09}, an accretion state
expected to be efficient in producing outflows and jets, and then to
correspond to the "radio-mode" of MBH feedback invoked in
semi-analytic studies and hydrodynamic simulations of galaxy formation
\citep[e.g.,][]{cro}.

\begin{figure*}[t!]
\hskip 3truecm
\resizebox{8cm}{!}
{\includegraphics{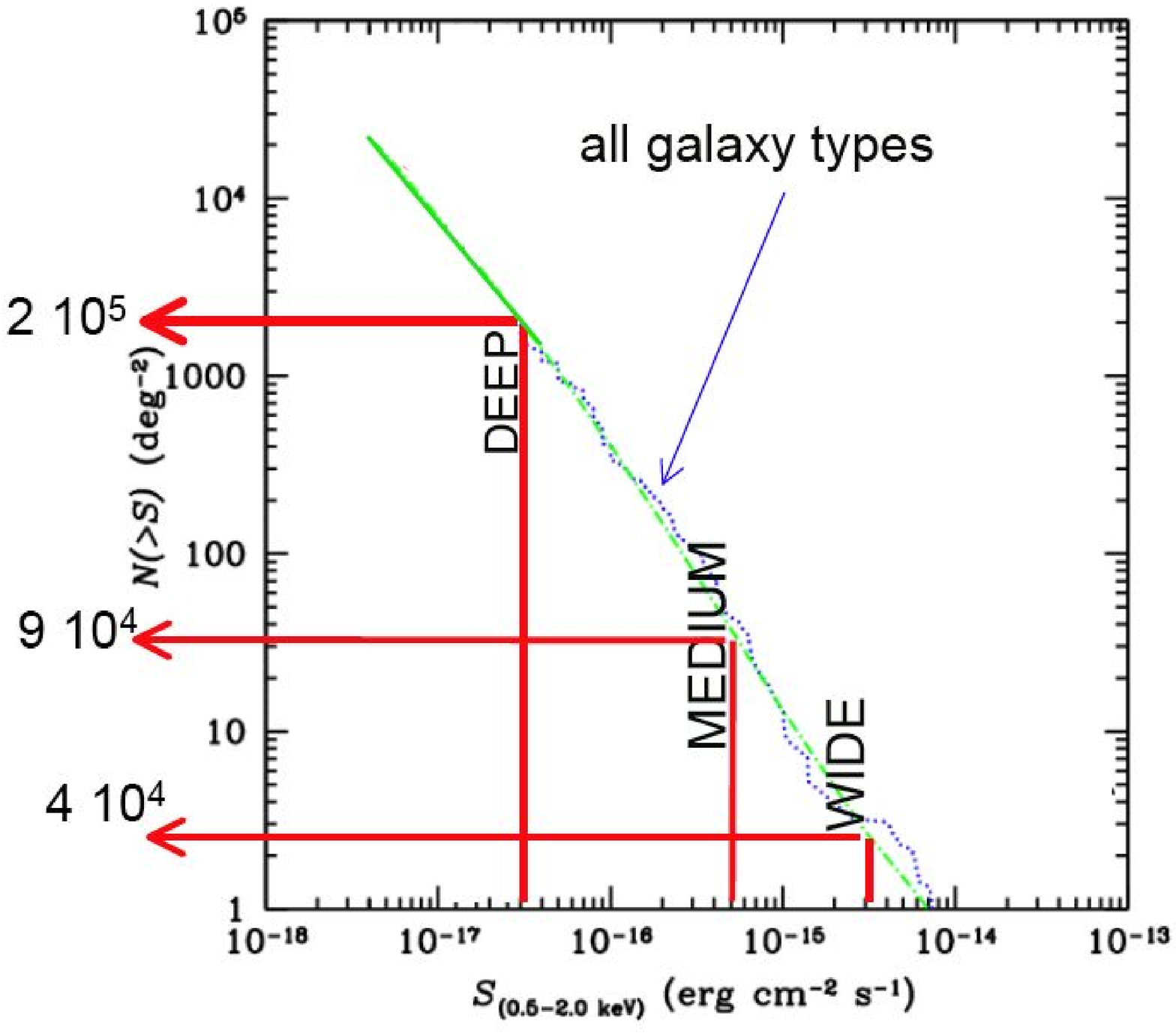}}
\caption{\footnotesize
The number of galaxies to be detected in the three $WFXT$
surveys (indicated by the arrows),
based on a recent estimate extrapolated to low
fluxes (green line) of the number $N(>S)$ of galaxies
with 0.5--2 keV
flux $S$ larger than the value on the x-axis (\citealt{tzav}). The green
line refers to both late and early types, with the galaxies in the sample
roughly equally divided between the two types.}

\label{f4}
\end{figure*}

\section{Beyond the local universe}\label{hiz}

Many surveys with different depths and fields of view have been
performed so far with $Chandra$ and $XMM-$Newton; for each of them
typically a sample of only $< \sim 100$ ETGs could be built, so that 
only few results could be obtained about the evolution with
redshift of ETGs.  The deepest study was conducted in the GOODS fields, where
40 ETGs divided in two redshift bins, of $z<0.5$ and $0.5<z<1.2$,
showed luminosity evolution, by which ETGs were brighter in the past;
this could be due to passive evolution of LMXBs
\citep[][Sect.~\ref{dscr}]{ptak}.  In the ECDF-S regions, 539
optically selected ETGs with $0.1<z<0.7$ and $R<24$ corresponded to
the detection of 13 luminous ETGs plus 32 AGNs, and the
characterization via the stacking procedure of the others
\citep{leh}. When divided in four $z$-bins from $z=0.25$ to $z=0.66$,
and two luminosity bins separated at $L_B\sim 10^{10}L_{B\odot}$, the
optically faint samples seem to show an increase in $L_X$ with $z$,
while the brighter ones keep within the range of values observed
locally, as due to a long-lasting ($\sim 6$ Gyr) balance between
heating and cooling of the hot gas coronae.  The wide area ($\sim 30$
deg$^2$) ChaMP survey based on archival $Chandra$ fields
\citep{kimchamp} for a sample of $<\sim 100$ ETGs at $0.01<z<0.3$
finds the minimum X-ray--to--optical ratio (likely the baseline
contributed by LMXBs) to be constant with redshift.  In the wide area (9.3
deg$^2$) XBo\"otes survey studied with a mosaic of 5 ks
pointings, the hardness ratio of 2968 stacked ETGs evolves from
$z=0.2$ to $z=0.4$, i.e., the average spectrum becomes harder with
increasing $z$, which could be due to an increasing AGN contribution
\citep{wat09}.  A collection of data from the $Chandra$ Deep Fields to
XBo\"otes, the shallowest survey, produced a sample of 101 ETGs up to
$z\sim 1.4$, that show no significant luminosity evolution when
divided in two $z$-bins centered at 0.17 and 0.67 \citep{tzav}.
Overall these investigations are heterogeneous, based on different
selection criteria, and plagued by the limited numbers of ETGs in the
samples, so that the results can be considered only preliminary.

\section{WFXT}\label{wfxt}

As discussed in the previous Sections, X-ray information about ETGs in
the local universe (distance $<100$ Mpc) is mostly based on a number of $<\sim
200$ galaxies that benefitted of pointed observations with $Chandra$ and
$XMM$-Newton.  Beyond the local universe, only samples with
$<100$ objects could be built, with very limited information on them;
future extensions of the surveys performed so far are not likely to
produce substantial improvements.  Statistical studies, as the
building of the ETG's LF and the search for its possible evolution, or the
study of the dependence of
the X-ray emission on different kinds of environment, require far
larger samples.  $WFXT$ is designed to produce a dramatic advance over
existing or planned missions in combined solid angle and sensitivity,
keeping a good angular resolution of $5^{\prime\prime}$ (see, e.g.,
Rosati in these proceedings); the energy band (0.4--7 keV) is
sensitive to both the soft hot gaseous emission and the hard stellar/AGN
contribution. With the three surveys (wide of 20000 deg$^2$, medium of
3000 deg$^2$ and deep of 100 deg$^2$) $WFXT$ could drastically
increase the number of detected ETGs and revolutionize the field
(see Fig.~\ref{f4}, based on the flux limits indicated by Rosati).  
For example, the deep survey is expected to
produce $\sim 10^3$ times the solid angle of the $Chandra$ Deep Fields
at the same sensitivity. An ETG with a (conservative) size of $\sim
20$ kpc will have an angular dimension of $10^{\prime\prime}$ at
$z=0.1$ and $5^{\prime\prime}$ at $z=0.3$, beyond which it will appear
as a pointlike source for $WFXT$.  Using flux limits for point sources, an
average ETG X-ray luminosity of $10^{41}$ erg s$^{-1}$ will be detected
out to $z=1$, 0.3 and 0.1 respectively in the deep, medium and wide
surveys.  The combination of the X-ray data with 
photometric and spectroscopic information at other wavelengths like
those provided by current and planned surveys (as 2MASS, SDSS, Galex,
LSST, BOSS, ...)  should give 
distances and the main galactic parameters.

The wide survey then could detect $\sim$few$10^4$ ETGs mostly within
$z=0.1$, and allow to build the first really large sample of ETGs in
the local universe. More than $\sim 10^3$ objects could be studied with
enough detail to measure gas properties, and distinguish stellar and nuclear
luminosities; angular resolution could enable to detect sharp features
in the hot gas as shocks, holes, rims.  This could make up a baseline
for medium/high $z$ studies.  Sample questions to be tackled with a
large database include: how is feedback working at all galactic
luminosities?  with what duty cycle?  is the large dispersion in hot
gas content (Fig.~\ref{f2}) related to nuclear activity, galaxy
structure, or environment?

At $z>0.1$ (a lookback time larger than 1.3 Gyr for standard
cosmological parameters), instead, the three main components (LMXBs,
hot gas and nuclei) should be revealed mostly from their integrated
contribution to the X-ray spectra, and their evolution could be
studied.  For example, the LMXB's contribution, determined at $z=0$ as
described in Sect.~\ref{dscr}, at $z>0$ should be higher than in local
ETGs, depending on epoch of major star formation (Sect.~\ref{dscr}).
The evolution of hot gas and nuclear activity (respectively
contributing to the soft and hard bands) should give important
insights on the feedback process, revealing for example whether the
hot gas content and temperature evolve with time, and the nuclear
luminosity increases.  In the deep survey there will be detections of
ETGs out to $z\sim 1$, to study the transition of accretion in the
radio mode and its evolution in this state (Sect.~\ref{nuc}).

\bibliographystyle{aa}

\end{document}